\documentclass[runningheads]{llncs}
\usepackage[T1]{fontenc}
\usepackage{etoolbox}
\usepackage{bm}

\usepackage{amsthm}%liwen
\usepackage{algorithm}%liwen
\usepackage{algorithmic}%liwen
\usepackage{subcaption}%dongang
\usepackage{xspace}
\usepackage{url}

\newcommand{\MyMapTemplatePrefixc}[4]{\expandafter#1\csname#3#4\endcsname{#2{#4}}} % it remembles a template: \#3#4 --> #2{#4}
\forcsvlist{\MyMapTemplatePrefixc {\def} {\mathcal}{c}} {A,B,C,D,E,F,G,H,I,J,K,L,M,N,O,P,Q,R,S,T,U,V,W,X,Y,Z}  % e.g., \cA --> \mathcal{A}

\newcommand{\MyMapTemplatePrefixtb}[5]{\expandafter#1\csname#4#5\endcsname{#2{#3{#5}}}} % it remembles a template: \#3#4 --> #2{#4}
\forcsvlist{\MyMapTemplatePrefixtb {\def} {\tilde}{\mathbf}{t}} {A,B,C,D,E,F,G,H,I,J,K,L,M,N,O,P,Q,R,S,T,U,V,W,X,Y,Z}  % e.g., \tA --> \tilde{A}
\forcsvlist{\MyMapTemplatePrefixtb {\def} {\tilde}{\mathbf}{t}} {0,1,a,b,c,d,e,f,g,h,i,j,k,l,m,n,o,p,q,r,s,u,v,w,x,y,z}  % e.g., \ta --> \tilde{a}

\newcommand{\MyMapTemplateNoPrefix}[3]{\expandafter#1\csname#3\endcsname{#2{#3}}}
\forcsvlist{\MyMapTemplateNoPrefix {\def} {\mathbf} } {0,1,a,b,d,e, f, g, h, j, k, m, n, p, q, r, u, v, w, x, y, z} % e.g., \a --> \mathbf{a} except for c, v, o, i, l
\forcsvlist{\MyMapTemplateNoPrefix {\def} {\mathbf} } {A,B,C,D,E,F,G,H,I,J,K,L,M,N,O,P,Q,R,S,T,U,V,W,X,Y,Z}  % e.g., \A --> \mathcal{A}

\usepackage{booktabs}       % professional-quality tables
\usepackage{lipsum}		% Can be removed after putting your text content
\usepackage{doi}
\usepackage{multicol}
\usepackage{multirow}
\usepackage{tabularray}
\usepackage{amssymb}
\usepackage{pifont}% http://ctan.org/pkg/pifont
\usepackage{amsmath} 
\usepackage{amsfonts}
\usepackage{bbm}
\usepackage{rotating}

\newcommand{\cmark}{\ding{51}}%
\newcommand{\xmark}{\ding{55}}%

\usepackage{graphicx}
\begin{document}
\title{Symmetry Awareness Encoded Deep Learning Framework for Brain Imaging Analysis}
\titlerunning{Symmetry Awareness Encoding for Brain Imaging}


\author{Yang Ma\index{Ma, Yang} \inst{1,2} \and
Dongang Wang\index{Wang, Dongang}\inst{2,5} \and
Peilin Liu\index{Liu, Peilin}\inst{2,3} \and
Lynette Masters\index{Masters, Lynette}\inst{4} \and
Michael Barnett\index{Barnett, Michael}\inst{2,5} \and
Weidong Cai \index{Cai, Weidong}\inst{1} \and
Chenyu Wang \index{Wang, Chenyu}\inst{2,5}
}
\authorrunning{Y. Ma et al.}
\institute{School of Computer Science, University of Sydney, NSW 2008, Australia \and
Brain and Mind Centre, University of Sydney, NSW 2050, Australia \and
School of Mathematics and Statistics, University of Sydney, NSW 2050, Australia \and
I-MED Radiology Network, NSW 2050, Australia \and
Sydney Neuroimaging Analysis Centre, NSW 2050, Australia\\
\email{chenyu.wang@sydney.edu.au}}
\maketitle              % typeset the header of the contribution
\begin{abstract}
% The abstract should briefly summarize the contents of the paper in

The heterogeneity of neurological conditions, ranging from structural anomalies to functional impairments, presents a significant challenge in medical imaging analysis tasks. Moreover, the limited availability of well-annotated datasets constrains the development of robust analysis models. Against this backdrop, this study introduces a novel approach leveraging the inherent anatomical symmetrical features of the human brain to enhance the subsequent detection and segmentation analysis for brain diseases. A novel Symmetry-Aware Cross-Attention (SACA) module is proposed to encode symmetrical features of left and right hemispheres, and a proxy task to detect symmetrical features as the Symmetry-Aware Head (SAH) is proposed, which guides the pretraining of the whole network on a vast 3D brain imaging dataset comprising both healthy and diseased brain images across various MRI and CT. 
% The pretraining framework is further augmented by a strategic combination of inpainting, symmetry, contrastive learning, and rotation tasks, each contributing uniquely to the model's robustness and generalizability. 
Through meticulous experimentation on downstream tasks, including both classification and segmentation for brain diseases, our model demonstrates superior performance over state-of-the-art methodologies, particularly highlighting the significance of symmetry-aware learning. Our findings advocate for the effectiveness of incorporating symmetry awareness into pretraining and set a new benchmark for medical imaging analysis, promising significant strides toward accurate and efficient diagnostic processes. Code is available at \href{https://github.com/bitMyron/sa-swin}{https://github.com/bitMyron/sa-swin}.

\keywords{Symmetry-Aware Cross-Attention (SACA) \and Self-Supervised Learning \and Neuroimaging Diagnosis }
\end{abstract}
\section{Introduction}

\begin{figure}
\centering
\includegraphics[width=0.7\textwidth]{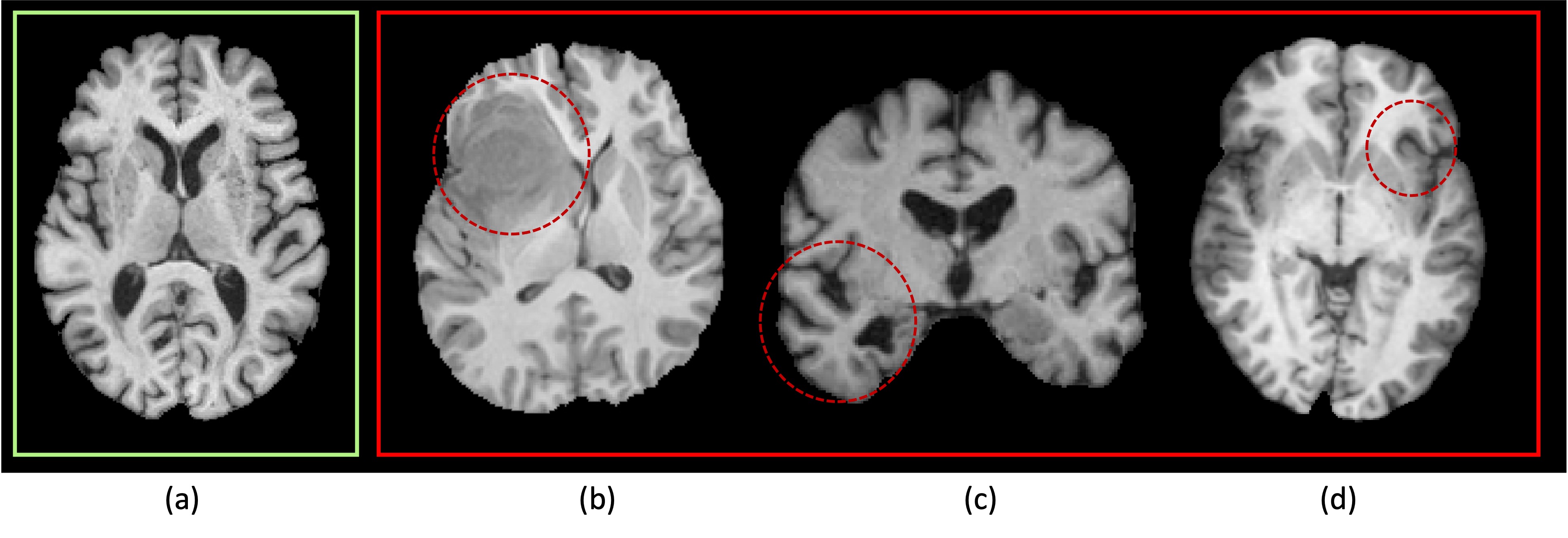}
\caption{Brain T1 images: (a) represents a healthy case, displaying symmetrical shape, structure, and image intensity. (b)-(d) depict patients with a brain tumor, Alzheimer's Disease, and focal epilepsy, respectively, and the noticeable structural asymmetry parts are circled out.} \label{example}
\end{figure}

In many neurological and psychiatric conditions, the symmetry nature of the left and right brain hemispheres plays a pivotal role in brain disease diagnosis and monitoring, often acting as a precursor to various disorders. The measurement of pathological structural and functional asymmetries could also be used in managing disease progression and the effectiveness of treatment~\cite{kuo2022structural}. 
The significance of these asymmetrical alterations is especially marked in diseases like Alzheimer's disease (AD)~\cite{illan2012bilateral} and schizophrenia~\cite{narr2007asymmetries}, where hemispheric symmetry is also considered as one of the most indicative biomarkers by radiologists. Figure~\ref{example} showcases instances of pathologies affecting the brain's symmetry.

Leveraging the clinical experience and identifying such asymmetrical features in neuroimaging analysis tasks can further enhance the efficacy of deep learning tools used in triage, annotation, and diagnosis. Certain approaches attempt to identify asymmetries by subtracting the L/R flipped image from the original and using the resulting subtraction as input for deep neural networks~\cite{herzog2021deep}. 
Other strategies involve segmenting the regions of interest on the original and flipped images separately, employing the pair of segmentation masks to compute an additional loss to improve model performance~\cite{hua2022symmetry}. However, hemispheric symmetrical features were only used for pre-processing or data augmentation rather than being encoded within the network structure, which limits the generalization ability to extend to the general neuroimaging analysis process. 

% Moreover, there has been a lack of attempts to apply symmetry understanding to 3D inputs, marking an unexplored area for advancing diagnostic accuracy.

% drawing inspiration from the diagnostic approaches of brain experts who frequently assess the brain's hemispheric symmetry for thorough disease analysis

To better leverage symmetrical features, we introduce an innovative Symmetry-Aware Cross-Attention (SACA) module, advancing the attention mechanism's application in neuroimaging analysis.
% , particularly for models trained on expansive datasets
Our module enhances the transformer blocks' capabilities by implementing cross-attention mechanisms between the original input and its symmetrical counterpart, compelling the network to focus on symmetrical comparisons and semantic feature encoding. This approach mimics the expert diagnostic strategy, leveraging self-supervised and supervised training processes to improve the model's understanding of brain anatomy and its inherent symmetries.

%~\cite{xu2023levit,zhou2021nnformer,wang2023swinmm}

% This module is based on attention mechanisms, which process information across the entire input sequences and have been shown to meet and then exceed the capabilities of traditional convolutional architectures such as ResNet~\cite{he2016deep} and U-Net~\cite{cciccek20163d}. The introduction of transformer blocks has heralded substantial advancements 

The SACA module can be pretrained with a substantial amount of data based on a self-supervised learning framework to be further used for downstream tasks, including disease diagnosis, brain structure, and lesion segmentation. The advancement of self-supervised learning has been illustrated in previous work~\cite{atito2021sit}, which involve generating pre-text tasks, including inpainting~\cite{haghighi2021transferable}, rotation~\cite{gidaris2018unsupervised} and contrastive coding~\cite{azizi2021big}, to embed intrinsic features. Besides these tasks widely used in self-supervised learning methods, we introduce a Symmetry-Aware Head (SAH) to enhance the pretraining process by focusing on the contrastive symmetrical features of the left and right hemispheres. Since they are entangled with the patient status (i.e., healthy or disease), the hemispheric symmetrical features are utilized in computing the symmetry-aware loss. 
% The advancement of self-supervised learning has facilitated significant progress in using unlabeled data effectively. This representation learning method, as discussed in works like~\cite{dosovitskiy2020image, atito2021sit, dai2021up}, involves generating pre-text tasks to construct embedding spaces for features. Popular tasks include impainting for predicting missing image patches, as detailed in~\cite{haghighi2021transferable, pathak2016context, zhou2021models}, predicting the rotation angle of randomly rotated images~\cite{gidaris2018unsupervised, taleb20203d}, and contrastive coding to promote patch similarity recognition from the same source for visual representation learning~\cite{chenempirical, oord2018representation, azizi2021big, chen2019self}. Our approach enhances traditional pre-training methodologies by incorporating a multifaceted strategy that leverages the Symmetry-Aware Cross-Attention (SACA) module to focus on symmetry attention alongside integrating a distinctive symmetry loss. This innovative combination not only adheres to established pre-training practices but significantly augments them by embedding disease-specific insights into the model. 

To fulfill the pretraining, we have compiled an extensive dataset comprising brain MRI and CT images from both healthy subjects (assumed to be symmetrical) and diseased subjects (assumed with asymmetrical features). Empirical experiments of our approach, through extensive pretraining on brain CT and MRI datasets and subsequent evaluations on various classification and segmentation tasks, underscore a notable improvement over existing baseline models. 

In general, our primary contributions are outlined as follows:

\begin{itemize}
\renewcommand{\labelitemi}{$\bullet$} 
% \item This work represents the inaugural exploration of the brain's intrinsic symmetry for analytical purposes.
\item We introduce a novel Symmetry-Aware Cross-Attention (SACA) module, which leverages a cross-attention mechanism to analyze the relationship between an image and its symmetrical counterpart, facilitating a deeper understanding of brain anatomy and its inherent symmetries. 
\item The proposed network is pretrained on a symmetry-aware self-supervised process and can be further applied in real-world clinical datasets and analysis tasks, including classification and segmentation for multiple modalities.
\item A vast dataset spanning MRI and CT modalities has been curated, and our approach has undergone extensive testing and demonstrated state-of-the-art performance on diagnosis and segmentation tasks.
\end{itemize}

% This confirms the benefit of incorporating symmetry awareness into the analysis of brain imaging. our research highlights the unexplored potential of symmetry-aware algorithms in the context of brain disease recognition, marking a significant step forward in the field of medical image analysis and paving the way for more accurate and efficient diagnostic procedures.

\section{Methodology}

In this study, we leverage a substantial corpus of 3D brain imaging data, denoted as $\mathcal{I}^0 = \{\I_n\}$. Each image is associated with a binary label $\mathcal{Y}^0=\{0,1\}$ indicating the health status of the subject (healthy or diseased), where $y_n^0 \in \mathcal{Y}^0$. Our objective is to train a model 
% $f_0$, parameterized by $\boldsymbol{\theta}^0$, 
to encapsulate structural insights into the human brain. The comprehensive pipeline
% , incorporating preprocessing and symmetry-aware pretraining, 
is depicted in Figure~\ref{pipeline}.

% This model is intended to particularly exploit the hemispheric symmetry inherent in brain anatomy, which could significantly enhance the performance of downstream tasks by hypothesis. 

\subsection{Preprocssing}

A midway registration pipeline was introduced to standardize the input image to position the brain within a standardized 3D coordinate framework, ensuring that flipping the image volume along its sagittal axis could generally map the left and right cerebral hemispheres correctly. 
Given an original input image $\I$, we first generate its mirror image $\I'$ by flipping $\I$ across the vertical plane of the image. We then obtain the affine transformation $\mathcal{T}: {\I} \rightarrow {\I}'$ with the widely used linear registration tool FMRIB's Linear Image Registration Tool (FLIRT)~\cite{jenkinson2002improved}. Based on the generated transformation matrix, the midway points of all transformation operations can be further calculated and formed the midway transformation $\mathcal{T}_{1/2}$. The midway transformation is applied to $\I$, resulting in $\Bar{\I}$, whose sagittal plane is perfectly aligned with respect to the vertical plane of the image volume. By the midway registration, the simple flip of the input image ensures the mirroring of the left and right hemispheres, assisting the model in learning and utilizing the inherent hemispheric symmetry of the human brain. 
 
After the midway registration, random sub-volumes, denoted as $\mathcal{X} = \{\x_n\}$, are extracted as inputs to networks from the collection of all aligned images $\mathcal{\Bar{I}} = \{\Bar{\I}_n\}$.
% to enhance the pretraining process through symmetry exploitation from the registered 3D brain image. 
Since the input images are standardized by the midway registration, for each patch $\x \in \mathbb{R}^{H \times W \times D}$, the mirrored counterpart $\x'$ with respect to the sagittal plane $\Phi_s$ can be identified and can be flipped to obtain $\tx$. Note that the generated $\tx$ is expected to closely resemble the original patch $\x$, barring minor structural discrepancies inherent to the intrinsic hemispheric symmetry of the human brain. Both original patch $\x$'s and their flipped symmetrical counterparts $\tx$'s are used as the input to the encoder network as sub-volumes. 
% This methodology leverages the symmetrical nature of brain anatomy to enrich the model's learning process, facilitating a nuanced understanding of cerebral structures.

\begin{figure}[t]
\centering
\includegraphics[width=0.96\textwidth]{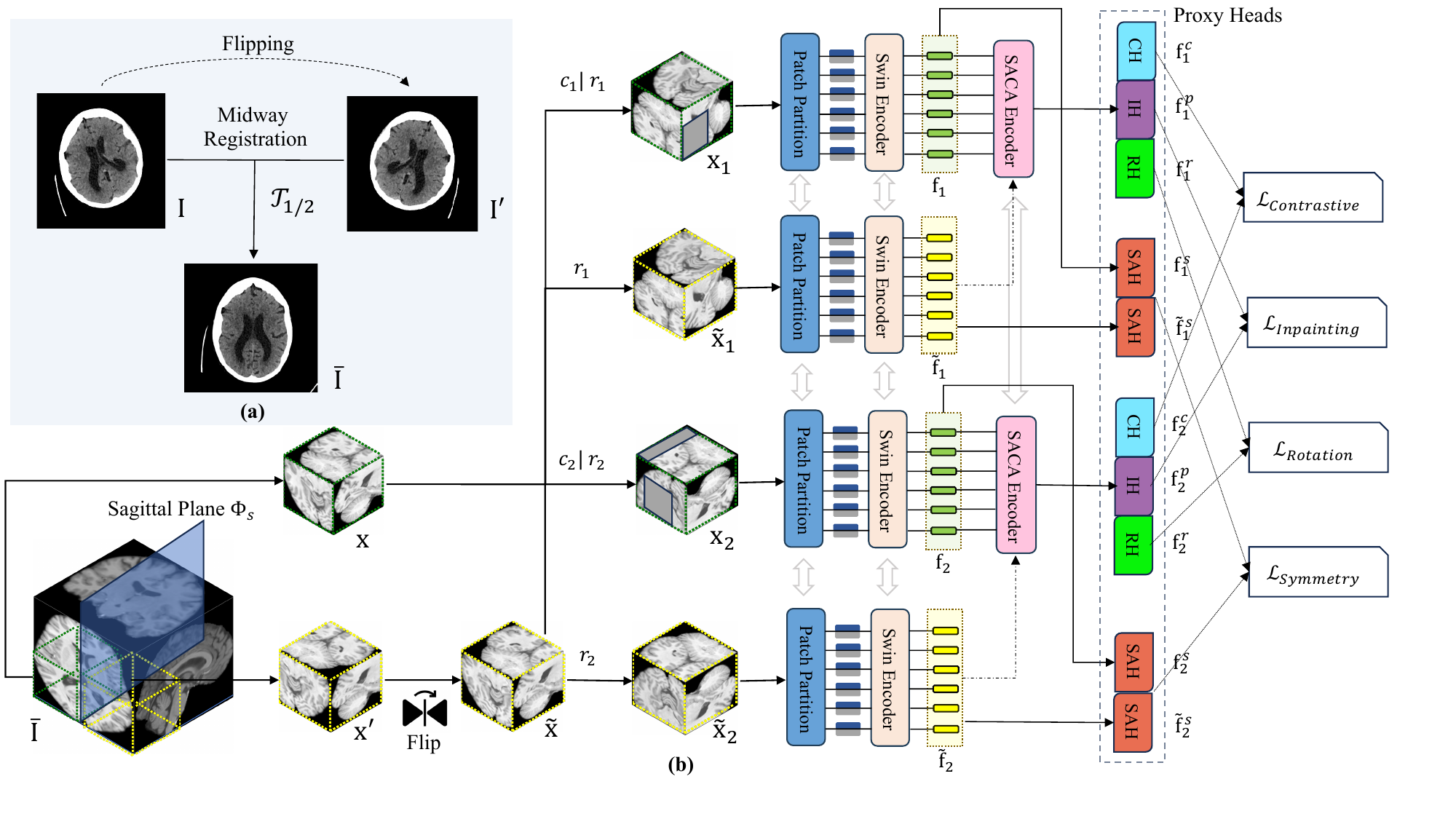}
\caption{(a) The process of midway registration to align input images; (b) The framework applied to input sub-volumes and their sagittal symmetrical counterparts. This framework is augmented by the Symmetry-Aware Cross-Attention (SACA) module and is optimized through a composite loss function that integrates inpainting, rotation, contrastive, and symmetry losses. 
% This comprehensive approach ensures the model's adeptness at capturing and utilizing symmetrical features within the data, significantly enhancing its learning efficacy. 
} \label{pipeline}
\end{figure}

As illustrated in Figure~\ref{pipeline}, two distinct rotations $R=\{r_1, r_2\}$ are conducted on both sub-volumes and two random inner cutouts $C=\{c_1, c_2\}$ are conducted on original sub-volume $\x$, resulting in augmented sub-volumes $\{\x_1, \x_2, \tx_1, \tx_2\}$.
% The augmentation of image sub-volume $x$ is performed through two distinct sets of transformations: random inner cutouts ${c_1, c_2}$ and rotations ${r_1, r_2}$, yielding augmented sub-volumes $x_{v1}$ and $x_{v2}$. To ensure consistent orientation with these augmented sub-volumes, the flipped mirrored counterpart $x_s^f$ undergoes identical rotations ${r_1, r_2}$, resulting in $x_{v1}^f$ and $x_{v2}^f$. Each of these resultant sub-volumes, denoted as $x_v^\circ$, is then processed through the designated encoder network $f_{\theta^0}$, which utilizes shared parameters $\theta^0$.
The augmented sub-volumes are segmented into patches of the size $(h, w, d)$, ending up with a sequence of the length $\frac{H}{h} \times \frac{W}{w} \times \frac{D}{d}$. These patches are processed through an embedding network and subsequently passed into a 3D Sliding-window (Swin) Transformer architecture~\cite{liu2021swin}, which consists of four stages and eight transformer layers. 
This results in a reduced sequence of the length $\frac{H}{16h} \times \frac{W}{16w} \times \frac{D}{16d}$ with the embedding feature dimension $768$.

\subsection{Symmetry-Aware Cross-Attention (SACA) Module}

\begin{figure}[t]
\centering
\includegraphics[width=0.92\textwidth]{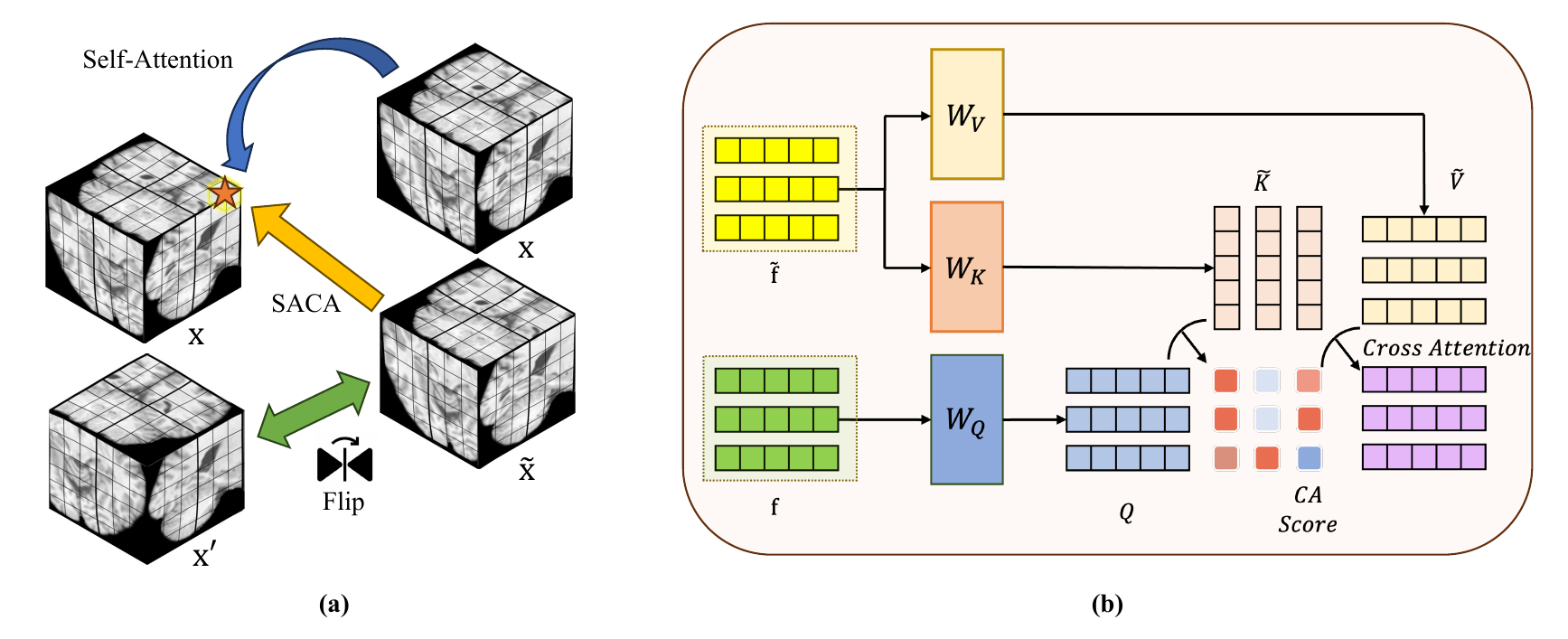}
\caption{Symmetry-Aware Cross-Attention (SACA) Module: (a) The computation of attention for each token within the SACA module; (b) Cross-attention calculation of SACA module structure.} \label{saca}
\end{figure}

Following the Swin Transformer Encoder~\cite{hatamizadeh2021swin}, we introduce a Symmetry-Aware Cross-Attention (SACA) module to foster deep symmetrical feature alignment and referencing. The SACA module processes input embedding sequences $\f$, applying self-attention mechanisms to $\f$ and cross-attention to the flipped symmetrical counterpart $\tf$. This dual attention strategy, as depicted in Figure~\ref{saca}, is intended to exchange information between the raw sub-volume and the flipped symmetrical counterpart. 
% central to the module's design, enabling it to effectively discern and leverage symmetrical patterns within the data, thus enhancing the model's analytical capabilities in recognizing and interpreting complex features in medical imaging.

The cross-attention mechanism within this module is represented as follows:

\begin{equation}
\operatorname{SACA}(\f, \tf) = \operatorname{Softmax}\left(\frac{Q\Tilde{K}^T}{\sqrt{D}}\right)\Tilde{V}, \text{where  } \Bigg\{
\begin{aligned}
Q &= \f W_Q \\
\Tilde{K} &= \tf W_K \\
\Tilde{V} &= \tf W_V
\end{aligned}
\ \ \ .
\end{equation}
$Q$ represents the queries generated from the features of the original patch $\f$, and $\Tilde{K}$ and $\Tilde{V}$ are keys and values generated from the counterpart patch $\tf$, with $D$ denoting the dimensionality of the query and key vectors. The matrices $W$ are learnable weights for the query, key, and value transformations within the SACA module. With our novel SACA module, the contrastive information from the original patch and its flipped symmetrical counterpart could be utilized, and the level of symmetry of the patch could be learned along the process, enabling the network to recognize symmetrical features of input brain images.

% \begin{equation}
% \operatorname{Attention}(Q,K,V) = \operatorname{Softmax}\left(\frac{\operatorname{Tril}(QK^T)}{\sqrt{D}}\right)V,
% \end{equation}
% \begin{equation}
% \begin{aligned}
% Q &= \operatorname{concat}(XW_Q, \Tilde{X}\Tilde{W}_Q), \\
% K &= \operatorname{concat}(XW_K, \Tilde{X}\Tilde{W}_K), \\
% V &= \operatorname{concat}(XW_V, \Tilde{X}\Tilde{W}_V), 
% \end{aligned}
% \end{equation}
% where $Q$, $K$, and $V$ represent the queries, keys, and values, respectively, with $D$ denoting the dimensionality of the query and key vectors. The matrices $W$ and $\Tilde{W}$ are learnable weights for the query, key, and value transformations within the SACA module. $\operatorname{Tril}$ denotes a lower-triangular mask on a matrix in $\mathbb{R}^{2D \times 2D}$ in the form of  
% \begin{equation}
%     \operatorname{Tril}(A) = 
% \begin{pmatrix}
% 1 & 0 & \cdots & 0 \\
% 1 & 1 & \ddots & \vdots \\
% \vdots & \ddots & \ddots & 0 \\
% 1 & \cdots & 1 & 1 \\
% \end{pmatrix} \odot A
% \end{equation}
% where $\odot$ denotes the point-wise matrix product.
% The $\operatorname{Tril}$ function ensures the attention is directionally biased, promoting an ordered and structured integration of symmetric features. This mechanism underpins the module's capability to effectively align and reference deep symmetric features, enhancing the model's sensitivity to anatomical symmetry in brain images.

% Upon processing through the symmetry-aware cross-attention (SACA) module, the features are directed to three distinct proxy heads, each responsible for a specific pretraining task.

\subsection{Symmetry-Aware Pretraining Proxies}

Upon obtaining the encoded features from the SACA-based encoder, they are 
directed to four distinct proxy heads, each responsible for a specific pretraining task. The proxies include one symmetry-aware head that identifies the symmetrical features of the input sub-volumes and three heads that deal with intrinsic features of the input sub-volume (i.e., inpainting, rotation, and contrastive coding) as shown in Figure~\ref{pipeline}.

The Symmetry-Aware Head (SAH) consists of a max pooling layer followed by a multi-layer perceptron (MLP), which distills the features for an original input patch and the symmetrical counterpart as $\f^s$ and $\tf^s$. These features serve as the foundation of computing the conditional symmetry loss, expressed as:

\begin{equation}
\mathcal{L}_{Symmetry}= -\log\frac{\sum \exp(\operatorname{sim}(\f^s, \tf^s)/ \tau ) \cdot y^0  }{\sum \exp(\operatorname{sim}(\f^s, \tf^s)/ \tau )}, 
\end{equation}
% \begin{equation}
% \operatorname{sim}(\mathbf{a},\mathbf{b}) = \frac{\mathbf{a}\cdot \mathbf{b}}{|\mathbf{a}|\cdot |\mathbf{b}|}
%   \label{eq2}
% \end{equation}
% where, $y_{si} \in \mathcal{Y}s$ represents the SAH output, and $y_{si}^f \in \mathcal{Y}_s^f$ is its symmetrical counterpart. 
where $y^0$ indicates the subject's health status, with 1 for healthy control and 0 for patients. The parameter $\tau$ is a temperature factor adjusting the logits' scaling. The cosine similarity $\operatorname{sim}(\mathbf{a},\mathbf{b})$ is used to measure the similarity between embedded features. 
The core premise of this formulation is that for subjects with healthy brains, a given patch and its symmetrical counterpart should exhibit similar visual and structural characteristics, whereas deviations and asymmetries are expected in the presence of pathological conditions, reflecting the brain's asymmetrical response to disease.

% , and $\epsilon$ is a small constant introduced to prevent division by zero.

The heads for intrinsic feature awareness follow~\cite{tang2022self}, including the Inpainting Head (IH), defined as a decoder to reconstruct the occluded voxels; the Rotation Head (RH), defined by an MLP classifier to identify the rotation angles; and the Contrastive Coding Head (CH), defined by two MLP layers to distinguish the present sub-volumes from other sub-volumes. 
% of VAE~\cite{baid2021rsna}
The losses are defined as:
\begin{equation}
\begin{aligned}
&\mathcal{L}_{Inpainting}  = \left \| \f^p - \x \right \|_1 \\
&\mathcal{L}_{Rotation}    = - r\log(\operatorname{Softmax}(\f^r)), \\
&\mathcal{L}_{Contrastive} = -\log\frac{\exp(\operatorname{sim}(\f^c_i, \f^c_j)/ \tau )}{\sum \mathbbm{1}_{k\neq i} \exp(\operatorname{sim}(\f_i^c, \f_k^c)/ \tau) },
\end{aligned}
\end{equation}
where $\f^p, \f^r, \f^c$ represent the concatenated output from each head, and $\x$ is the input sub-volume. 
Through comprehensive grid search experimentation, the four losses are configured to be evenly summed up for pretraining, balancing the contribution of each task effectively.
The pretrained symmetry-aware network can be further applied for downstream tasks, as shown in Section \ref{sec:exp}.

\section{Experiments and Results}\label{sec:exp}
% \vspace{-0.1em}

We evaluated our proposed approach by pretraining on large neuroimaging datasets and downstream classification and segmentation tasks.
% \vspace{-0.1em}

\subsection{Datasets and Downstream Tasks}
% \textbf{\textit{Pretraining Datasets:}} 
We curated two large datasets for brain image pretraining using head CT and brain MRI images separately. The MRI dataset comprises 3D T1 images of 3,509 healthy adults and 2,917 patients sourced from a variety of publicly available datasets~\cite{biswal2010toward,zuo2014open,rowland2004information,nooner2012nki,lamontagne2019oasis,jack2008alzheimer,baid2021rsna,jack2008alzheimer} (more details in supplementary). The CT dataset is derived from our proprietary in-house collection, including 2025 images from healthy individuals and 4693 images from patients afflicted with a range of brain diseases, with images from three brain diseases (i.e., hemorrhage, fracture, midline shift) reserved for downstream tasks.

%, including 867 images from the 1000 Connectomes Project~\cite{biswal2010toward}, 639 images from the Consortium for Reliability and Reproducibility (CoRR)~\cite{zuo2014open}, 560 images from the Information eXtraction from Images (IXI) database~\cite{rowland2004information}, 342 images from the Nathan Kline Institute Rockland Sample (NKI)~\cite{nooner2012nki}, 603 images from the Open Access Series of Imaging Studies 3 (OASIS-3)~\cite{lamontagne2019oasis}, and 498 images from the Alzheimer’s Disease Neuroimaging Initiative (ADNI)~\cite{jack2008alzheimer}. For cases presenting with brain disease, the dataset incorporates 1251 FLAIR images from the Brain Tumor Segmentation Challenge 2021 (BraTS2021)~\cite{baid2021rsna} and an additional 1666 images from ADNI~\cite{jack2008alzheimer}.
% , thus encompassing a wide spectrum of brain conditions, from prominent lesions and tumors to subtle structural changes and dysfunctions. 

% This comprehensive and meticulously compiled dataset serves as the foundation for our pretraining regimen, facilitating the development of robust models capable of discerning and analyzing complex brain image data.

% \noindent
% \textbf{\textit{Downstream Tasks:}} 
For MRI-based evaluations, we assess the zero-shot and few-shot classification capabilities of our model across several disease identification tasks, including Focal Epilepsy~\cite{ds004199:1.0.5}, ADHD~\cite{ds002424:1.2.0}, and Schizophrenia~\cite{soler2022brain}. 
Furthermore, we explore the model's transfer learning proficiency in segmentation tasks using images with different modalities from the MRI pretraining. This includes the BraTS2021 dataset~\cite{baid2021rsna} and the MSSEG2016 dataset~\cite{commowick2021multiple}. 

For CT imaging, we conducted few-shot training using reserved in-house data of three brain diseases, including intracranial hemorrhage, bone fracture, and midline shift, and evaluated the performance on the public dataset CQ500~\cite{chilamkurthy2018deep} to demonstrate the model's versatility across varied clinical conditions.

% \noindent
% \textbf{\textit{Baselines and Metrics:}} 
\subsection{Implementation Details}
% For our evaluation, we benchmark against the Swin MLP for classification tasks and the Swin UNETR for segmentation tasks as described by~\cite{tang2022self}. To ensure a fair comparison, we pre-train the referenced models using our assembled dataset. For the classification domain, we adopt Precision-Recall Area Under the Curve (PR-AUC), Receiver Operating Characteristic Area Under the Curve (ROC-AUC), Accuracy, and Maximum F1 Score as our evaluation metrics. In the context of segmentation tasks, we measure performance using the Dice Similarity Coefficient (DSC) metric, as outlined by~\cite{zou2004statistical}. 

% \noindent
% \textbf{\textit{Implementation Details:}} 
The proposed model undergoes pretraining with distinct settings for image classification and segmentation tasks. Specifically, the whole image classification task pretraining is conducted with a batch size of 2 and images resized to $160\times160\times160$, and a batch size of 4 and sub-volume dimensions of $96\times96\times96$ was used for segmentation-related pretraining. 
% The pretraining process employs a Hounsfield Unit (HU) range of $[-1000, 1000]$, which is normalized across datasets. For downstream MRI tasks, the HU values are clipped to the range of $[-175, 250]$, whereas for CT tasks, the clipping range is adjusted to $[-100, 300]$. 
The pretraining phase is executed with a learning rate of $6\times10^{-6}$ over 300 epochs. For downstream tasks, the learning rate is set to $1\times10^{-5}$, spanning 100 epochs. All models are trained utilizing NVIDIA Tesla V100 GPUs.
% , ensuring high computational efficiency and performance.

The pretrained models are further experimented with downstream tasks under a 5-fold cross-validation approach across all experiments, with the exception of the multiple sclerosis lesion segmentation on MSSEG, which has a predefined validation set ~\cite{ma2022multiple}. The average performance scores, including Area under the ROC Curve (AUC), F1-measure, and Dice Similarity Coefficient (DSC), are reported.

\subsection{Results}

The classification outcomes are delineated in Tables~\ref{tab:table1} and~\ref{tab:table2}. In the supervised fine-tuning scenario (checked under ``+SFT''), our model consistently surpassed the performance of the Swin MLP~\cite{tang2022self} across all evaluated downstream tasks, especially in some diseases that display significant asymmetrical features, such as ADHD and Midline Shift. 
Besides, our method has achieved comparable performance in CT disease diagnosis as that of~\cite{chilamkurthy2018development}, but only hundreds of labeled cases were used for fine-tuning, much less than in~\cite{chilamkurthy2018development} where 313,318 scans were collected and manually labeled. This underscores the significant advantage of incorporating symmetrical awareness during the pretraining phase. 

% \vspace{-2em}
\begin{table}[]
\centering
\caption{Benchmark methods on the MRI T1-only classification tasks.}
\label{tab:table1}
\begin{tabular}{lccccccccc}
\hline
\multirow{3}{*}{Experiments} &
   &
   &
   &
  \multicolumn{2}{c}{Focal Epilepsy} &
  \multicolumn{2}{c}{Schizophrenia} &
  \multicolumn{2}{c}{ADHD} \\ \cmidrule(lr){5-6} \cmidrule(lr){7-8} \cmidrule(l){9-10} 
 &
  +Pretrain&
  +SACA &
  +SFT &
  AUC &
  F1 &
  AUC &
  F1 &
  AUC &
  F1 \\ \hline
\multirow{3}{*}{Swin MLP~\cite{tang2022self}} &
 
  \xmark &
  \xmark &
  \cmark &
  0.678 &
  0.683 &
  0.656 &
  0.810 &
  0.685 &
  0.625 \\
  &
  \cmark &
  \xmark &
  \xmark &
  0.522 &
  0.651 &
  0.594 &
  0.749 &
  0.514 &
  0.609 \\
 &
  \cmark &
  \xmark &
  \cmark &
  0.713 &
  0.695 &
  0.627 &
  0.803 &
  0.642 &
  0.645 \\ \hline
\multirow{2}{*}{Ours} &
  \cmark &
  \cmark &
  \xmark &
  0.631 &
  0.683 &
  0.663 &
  0.760 &
  0.515 &
  0.625 \\
 &
  \cmark &
  \cmark &
  \cmark &
  \textbf{0.778} &
  \textbf{0.788} &
  \textbf{0.719} &
  \textbf{0.836} &
  \textbf{0.792} &
  \textbf{0.727} \\ \hline
\end{tabular}
\end{table}
% \vspace{-1em}
Furthermore, we assessed the impact of symmetrical awareness in a zero-shot learning context. Remarkably, our pretrained model exhibited superior zero-shot capabilities in almost all tasks. This indicates that our pretraining strategy effectively captures critical pathological features based on symmetry characteristics, facilitating generalization to novel tasks without requiring task-specific data.  
% Such results highlight the efficacy and versatility of our approach in enhancing brain MRI analysis with scant training resources, showcasing its potential for broad applications in medical imaging.

% Contrary to the baseline CT classification, which trained ResNet18 on 313,318 scans~\cite{chilamkurthy2018development}, our method employs only hundreds of cases for fine-tuning downstream tasks yet achieves comparable, if not superior, outcomes. This is particularly noteworthy given our use of a transformer-based network, renowned for its data-intensive training requirements. Our results, often surpassing the baseline across various tasks, underscore the efficiency and potential of our approach in leveraging limited data for high-performance medical imaging analysis. Our approach achieved remarkable improvements, particularly for tasks such as midline shift detection, which posed challenges for baseline methods. This is because the diagnosis of such conditions heavily depends on the observation of symmetry, underscoring the effectiveness of our method in leveraging this critical aspect.

% \vspace{-2em}
\begin{table}[]
\centering
\caption{Benchmark methods on the CT classification tasks evaluated on the CQ500 dataset~\cite{chilamkurthy2018development}. }
\label{tab:table2}
\begin{tabular}{lccccccccc}
\hline
\multirow{3}{*}{Experiments} &
   &
   &
   &
  \multicolumn{2}{c}{Hemorrhage} &
  \multicolumn{2}{c}{Fracture} &
  \multicolumn{2}{c}{Midline Shift} \\ \cmidrule(lr){5-6} \cmidrule(lr){7-8} \cmidrule(l){9-10} 
 &
  +Pretrain&
  +SACA &
  +SFT &
  AUC &
  F1 &
  AUC &
  F1 &
  AUC &
  F1 \\ \hline
  Qure.ai~\cite{chilamkurthy2018development} &
  \xmark &
  \xmark &
  \cmark &
  0.942 &
  0.761 &
  0.962 &
  0.508 &
  0.970 &
  0.704 \\
  \hline
\multirow{3}{*}{Swin MLP~\cite{tang2022self}} &
\xmark &
  \xmark &
  \cmark &
  0.872 &
  0.776 &
  0.918 &
  0.460 &
  0.926 &
  0.640 \\
  &
  \cmark &
  \xmark &
  \xmark &
  0.751 &
  0.685 &
  0.781 &
  0.353 &
  0.788 &
  0.413 \\
 &
  \cmark &
  \xmark &
  \cmark &
  0.911 &
  0.748 &
  0.952 &
  0.469 &
  0.942 &
  0.620 \\
   \hline
\multirow{2}{*}{Ours} &
  \cmark &
  \cmark &
  \xmark &
  0.793 &
  0.722 &
  0.803 &
  0.267 &
  0.866 &
  0.508 \\
 &
  \cmark &
  \cmark &
  \cmark &
  \textbf{0.958} &
  \textbf{0.781} &
  \textbf{0.969} &
  \textbf{0.513} &
  \textbf{0.986} &
  \textbf{0.728} \\ \hline
\end{tabular}
\end{table}

The assessment of models' proficiency in adapting to segmentation tasks is summarized in Table~\ref{tab:table3}, which manifests the superior performance of our method.
Additionally, we conducted an ablation study to discern the relative contribution of our proposed SACA module and SAH proxy. Both the SAH only pretraining process and the SACA module could improve the performance of segmentation steadily, and the combination of all pretraining proxies and the novel SACA module could further yield the most substantial benefits, underscoring the synergetic effect of combining these diverse learning strategies. 
This comprehensive evaluation demonstrates our method's superior capability in leveraging pretraining proxies to enhance model performance on segmentation tasks, highlighting its potential for advancing medical image analysis.

% \vspace{-2em}
\begin{table}[]
\centering
\caption{Benchmark methods and ablation study of our method on the MRI segmentation tasks. ``IRC'' represents commonly used pretrain heads for inpainting (I), rotation (R), and contrastive (C).}
\label{tab:table3}
\begin{tabular}{lcccccccccc}
\hline
\multirow{3}{*}{Experiments} &
   &
   &
    
  \multicolumn{4}{c}{BraTS~\cite{baid2021rsna}} &
  MSSEG~\cite{commowick2021multiple} \\ \cmidrule(lr){5-8} \cmidrule(lr){9-9}
 &
 +IRC&
 +SAH& 
 +SACA&
  $\operatorname{Dice}_{ET}$ &
  $\operatorname{Dice}_{TC}$ &
  $\operatorname{Dice}_{WT}$ &
  $\operatorname{Dice}_{Avg}$ &
  Dice  \\ \hline
    nnU-Net~\cite{isensee2021nnu} &
   \xmark &
   \xmark &
   \xmark&
  0.883 &
  0.927 &
  0.913 &
  0.908 &
  - \\
  %   SegResNet~\cite{myronenko20193d} &
  %   &&&
  % 0.883 &
  % 0.927 &
  % 0.913 &
  % 0.907 &
  % - \\
  TransBTS~\cite{wang2021transbts} &
  \xmark&
  \xmark &
  \xmark&
  0.868 &
  0.911 &
  0.898 &
  0.891 &
  -  \\ 
  % \hline
   SwinUNETR~\cite{hatamizadeh2021swin} &
  \xmark &
  \xmark &
  \xmark &
  0.891 &
  0.933 &
  0.917 &
  0.913 &
  0.594  \\ \hline
 \multirow{3}{*}{Ours} &
 \xmark &
  \cmark &
  \xmark &
  0.897 &
  0.935 &
  0.920 &
  0.917 &
  0.615  \\&
 
  \xmark &
  \xmark &
  \cmark &
  0.901 &
  0.937 &
  0.928 &
  0.922 &
  0.640  \\&

 \xmark &
  \cmark &
  \cmark &
  0.909 &
  0.940 &
  0.935 &
  0.928 &
  0.654  \\ 

  % &
  % Sym Only &
  % \cmark &
  % 0.909 &
  % 0.940 &
  % 0.935 &
  % 0.928 &
  % 0.654  \\ 

  &
  \cmark &
  \cmark &
  \cmark &
  \textbf{0.912} &
  \textbf{0.946} &
  \textbf{0.940} &
  \textbf{0.932} &
  \textbf{0.680}   \\ \hline
\end{tabular}
\end{table}

% \begin{table}[]
% \centering
% \caption{Benchmark pretraining methods on the MRI segmentation tasks.}
% \label{tab:table3}
% \begin{tabular}{lcccccccc}
% \hline
% \multirow{1}{*}{Experiments} &
%   \multicolumn{4}{c}{BraTS~\cite{baid2021rsna}} &
%   \multicolumn{2}{c}{MSSEG~\cite{commowick2021multiple}} \\ \cmidrule(lr){2-5} \cmidrule(l){6-7} 
%  &
%   $\operatorname{Dice}_{ET}$ &
%   $\operatorname{Dice}_{TC}$ &
%   $\operatorname{Dice}_{WT}$ &
%   $\operatorname{Dice}_{Avg}$ &
%   Dice &
%   F1 \\ \hline
%     nnU-Net~\cite{isensee2021nnu} &
%   0.883 &
%   0.927 &
%   0.913 &
%   0.908 &
%   - &
%   - \\
%     SegResNet~\cite{myronenko20193d} &
%   0.883 &
%   0.927 &
%   0.913 &
%   0.907 &
%   - &
%   - \\
%   TransBTS~\cite{wang2021transbts} &
%   0.868 &
%   0.911 &
%   0.898 &
%   0.891 &
%   0.571 &
%   0.389 \\
%    SwinUNETR~\cite{hatamizadeh2021swin} &
%   0.891 &
%   0.933 &
%   0.917 &
%   0.913 &
%   0.594 &
%   0.418 \\
%    \hline
%     SACA +  $\mathcal{L}_{Impainting}$ &
%   0.895 &
%   0.935 &
%   \textbf{0.923} &
%   0.917 &
%   0.603 &
%   0.437 \\
 
%  SACA +  $\mathcal{L}_{Rotation}$ &
%   0.892 &
%   0.931 &
%   0.918 &
%   0.914 &
%   0.597 &
%   0.420 \\

%   SACA +  $\mathcal{L}_{Contrastive}$ &
%   0.893 &
%   0.934 &
%   0.914 &
%   0.913 &
%   0.598 &
%   0.425 \\

%   SACA +  $\mathcal{L}_{Symmetry}$ &
%   0.893 &
%   0.937 &
%   0.919 &
%   0.916 &
%   0.601 &
%   0.435  \\ \hline

%   SACA +  $\mathcal{L}_{All}$ &
%   \textbf{0.897} &
%   \textbf{0.940} &
%   0.922 &
%   \textbf{0.920} &
%   \textbf{0.611} &
%   \textbf{0.449}  \\ \hline
% \end{tabular}
% \end{table}
% \vspace{-2em}

\section{Conclusion}
% \vspace{-1em}

Our study introduces an innovative framework utilizing 3D brain imaging data of various modalities, significantly enhancing neuroimage analysis by encoding symmetry-aware features in deep neural networks. The integration of the Symmetry-Aware Cross-Attention (SACA) module and the symmetry-aware pretraining proxy has shown remarkable improvements in classification and segmentation tasks, setting new benchmarks over existing models. This work highlights the critical role of anatomical symmetry in neuroimaging analysis and suggests a promising direction for developing more accurate, robust, and generalized AI models in healthcare leveraging intrinsic anatomical features.

\begin{credits}
\subsubsection{\ackname} The authors would like to express their gratitude to the BCR Scholarship for Computational Neuroscience and the Faculty of Engineering Research Stipend Scholarship at the University of Sydney for their financial support, which enabled this research and its contributions to medical imaging and healthcare.

\subsubsection{\discintname}
The authors have no competing interests to declare that are
relevant to the content of this article.
% state that the authors have no competing interests. Please place the
% statement with a bold run-in heading in small font size beneath the
% (optional) acknowledgments\footnote{If EquinOCS, our proceedings submission
% system, is used, then the disclaimer can be provided directly in the system.},
% for example: The authors have no competing interests to declare that are
% relevant to the content of this article. Or: Author A has received research
% grants from Company W. Author B has received a speaker honorarium from
% Company X and owns stock in Company Y. Author C is a member of committee Z.
\end{credits}


\bibliographystyle{splncs04}
\bibliography{Paper-2579}

\end{document}